\documentclass[preprint2]{aastex}

\shorttitle{SLACS gravitational lenses properties}
\shortauthors{Guimar\~aes \& Sodr\'e}

\slugcomment{}

\begin{document}

\title{Density profile, velocity anisotropy and line-of-sight external \\ convergence of SLACS gravitational lenses}

\author{Antonio C. C. Guimar\~aes and Laerte Sodr\'e Jr.}
\affil{Departamento de Astronomia, Universidade de S\~ao Paulo,  
Rua do Mat\~ao 1226,CEP 05508-090 S\~ao Paulo - SP, Brazil; aguimaraes@astro.iag.usp.br}

\begin{abstract}
Data from 58 strong-lensing events surveyed by the Sloan Lens ACS Survey
are used to estimate the projected galaxy mass inside their Einstein radii 
by two independent methods: stellar dynamics and strong gravitational lensing. 
We perform a joint analysis of these two estimates inside models with up to 
three degrees of freedom with respect to the lens density profile, stellar 
velocity anisotropy, and line-of-sight (LOS) external convergence, which
incorporates the effect of the large-scale structure on strong lensing.
A Bayesian analysis is employed to estimate the model parameters, 
evaluate their significance and compare models.
We find that the data favor Jaffe's light profile over Hernquist's, but that 
any particular choice between these two does not change the qualitative 
conclusions with respect to the features of the system that we investigate.
The density profile is compatible with an isothermal, being sightly steeper 
and having an uncertainty in the logarithmic slope of the order of 5\% 
in models that take into account a prior ignorance on anisotropy and 
external convergence.
We identify a considerable degeneracy between the density profile slope and 
the anisotropy parameter, which largely increases the uncertainties in the 
estimates of these parameters, but we encounter no evidence in favor of an 
anisotropic velocity distribution on average for the whole sample.
An LOS external convergence following a prior probability distribution 
given by cosmology has a small effect on the estimation of the  
lens density profile, but can increase the dispersion of its value by 
nearly 40\%.
\end{abstract}


\maketitle

\section{Introduction}

The observation of strong gravitational lensing events has allowed many 
studies about the mass, density profile, and structure of the galaxies that 
act as lenses, which could also have important implications for dark 
matter and cosmology studies --- see Kochanek's contribution in 
\cite{2006glsw.conf.....M} for a review in the field. 
Among strong lensing works, and of particular interest here, 
there are those of joint studies with stellar dynamics, which allow a 
determination of the density profile for individual lenses 
\citep{1999ApJ...516...18R,2002MNRAS.334..621T,2003ApJ...583..606K,
2002MNRAS.337L...6T,2002ApJ...575...87T,2003ApJ...599...70K,
2008MNRAS.384..987C} and for sets of lenses 
\citep{2006ApJ...649..599K,2007ApJ...667..176G,2009MNRAS.399...21B}.
It is of fundamental interest to this line of research to control for 
possible systematic effects, such as the influence of the large-scale 
structure (LSS) along the line-of-sight (LOS) and the role of other 
properties of the lens galaxy.

Several works have considered LOS effects on strong lensing studies.
\cite{1996ApJ...468...17B} investigated theoretically the effect of the 
LSS on strong-lensing events, finding that it can be 
significant, for example leading to incorrect conclusions about the distribution of matter in the lens. 
\cite{1997ApJ...482..604K} observed that external shear due to 
galaxies and clusters associated with the primary lens or along the LOS 
can be an important perturbation in individual lens models.
\cite{2004ApJ...611....1P} used ray tracing to investigate the effect of 
density inhomogeneities along the LOS of strong lenses and concluded that 
the effect of environment is negligible in general, but might be important in 
rare cases.
\cite{2005ApJ...635L...1W} noted that secondary matter along the LOS of 
strong lenses is strongly dependent on source redshift, being rare for sources 
with $z<1$, but that can lead in some circumstances to an overestimate of 
10\%--15\% of the primary lens mass if ignored.
Photometric and spectroscopic observations \citep{2000AJ....119.1078T,
2006ApJ...641..169M,2006ApJ...642...30F,2006ApJ...646...85W,
2007AJ....134..668A} discovered a significant LOS effect on some 
individual strong lens galaxies.
\cite{2007ApJ...660L..31M} observed that even in underdense local 
environments, the LOS contamination may give a considerable contribution 
to galaxy-scale strong lenses. 
Using ray-tracing thought the Millennium Simulation, 
\cite{2007MNRAS.382..121H} determined that strong-lensing LOSs 
are biased toward higher than average mean densities, contributing a few 
percent to the total surface density, 
and \cite{2009MNRAS.398.1298P} found that secondary matter along the LOS 
has a large effect on the strong-lensing optical depth and the cross section 
for cluster strong lensing. 
\cite{2008MNRAS.383L..40A} did not find an overdensity of 
photometric sources along the LOS of a limited sample of SLACS strong 
lenses in comparison with other Sloan digital Sky Survey (SDSS) massive 
early-type galaxies and interpreted that as evidence against a possible 
LOS contamination.
\cite{2009ApJ...690..670T} measured the overdensity of galaxies around SLACS 
lenses and observed that typical contributions from external mass 
distributions are of the order of few percent, but reaching 10\%--20\% in some cases.
\cite{2009ApJ...695.1233F} considered strong- and weak-lensing observations in 
the COSMOS survey and compared with simulations, finding that strong-lensed 
images with large angular separation were in the densest regions.

In this paper, we use two independent galaxy mass estimate methods, strong 
gravitational lensing, and stellar dynamics to examine the influence of the 
LSS in the LOS of the lenses and its effect on the determination of 
the lens density profile.
We use all the suitable events in the SLACS sample, considering realistic 
brightness functions for the lens galaxies, and incorporating our prior 
ignorance on their stellar velocity anisotropy.

In Section \ref{methods_data}, we describe how we calculate the mass of SLACS 
lenses using strong gravitational lensing and stellar dynamics, and 
our Bayesian statistical approach. 
In Section \ref{results} we show our results, and in 
Section \ref{conclusion} we present our discussion and conclusions.

\section{Data and Methods}
\label{methods_data}

The analysis in this paper is based on the comparison of galaxy masses 
calculated through two different methods: gravitational lensing and dynamical 
analysis. 
In Section 2.1, we present the data used in the analysis, collected
from the SLACS survey.

In Sections 2.2 and 2.3, we discuss the lensing and dynamical mass 
determinations, respectively. We will assume simple models for the galaxy 
mass distribution (e.g., spherical symmetry, power-law density distribution)
because they have few free parameters and allow to illustrate well the two 
methods. For a similar approach see \cite{2006ApJ...649..599K}. 

We want to examine whether the two mass estimates are indeed equivalent and/or
if there is evidence of systematic differences between them. 
Section 2.4 presents a Bayesian framework to analyze this problem.

\subsection{Data}
\label{data}

The selected set of galaxies is part of the Sloan Lens ACS Survey, SLACS
\citep{2006ApJ...638..703B}, which is a {\it Hubble Space Telescope (HST)} 
Snapshot imaging survey for strong gravitational galactic lenses. The 
candidates for the HST imaging were selected spectroscopically from the SDSS 
database and are a sub-sample of the SDSS Luminous Red Galaxy (LRG) sample.

We use data compiled from \cite{2006ApJ...649..599K}, 
\cite{2007ApJ...667..176G} and \cite{2008ApJ...682..964B}, constructing a 
sample of 58 strong gravitational lensing events where the lenses are 
isolated early-type galaxies (E+S0). 
Data from SLACS are especially suitable for joint strong lensing and dynamical 
analysis because they allow precise determination ($5\%$ error) of the 
Einstein radius for each lens galaxy in a relatively homogeneous sample of 
early-type galaxies. And, at the same time, SDSS has precise stellar velocity 
dispersion measurements ($6\%$ average error) for the lenses, as well as 
redshifts for lenses and background sources. 

For each lens system we are interested in the redshift of the
background lensed source $z_s$, the redshift of the lens $z_l$,
the average stellar velocity dispersion inside an aperture
$\sigma_{ap}$, the effective angular radius $\theta_{ef}$ and the 
Einstein angular radius $\theta_E$. 
The sample average values for these quantities are
$\left<z_l\right>=0.2$, $\left<z_s\right>=0.6$,
$\left<\sigma_{ap}\right>=250 km s^{-1}$,
$\left<\theta_{ef}\right>=2.2\arcsec$, and 
$\left< \theta_E \right>=1.2\arcsec$. 

The source and lens redshifts were determined from the SDSS spectra, and the 
stellar velocity dispersion corresponds to the light-weighted average inside 
the $3\arcsec$  diameter SDSS fiber.

\subsection{Lensing Mass}

The estimated projected mass inside the Einstein radius $R_E=\theta_E D_L$, 
is given by
\begin{equation}
M_E = \pi R_E^2 \Sigma_{cr}\;, 
\label{M_E}
\end{equation}
where 
\begin{equation}
\Sigma_{cr}=   \frac{c^2}{4\pi G}\frac{D_S}{D_L D_{LS}} \; ,
\end{equation}
and $D_{[L,S,LS]}$ is the angular-diameter distance of the
lens, source, and between lens and source, respectively. 
These distances are calculated assuming a redshift-distance relation
derived inside a chosen cosmological model that in the present
paper is a concordance $\Lambda$CDM model with $\Omega_m=0.3$,
$\Omega_{\Lambda}=0.7$.

The Einstein radii were determined from {\it HST} images using strong lensing 
modeling of the lenses and reconstruction of the unlensed sources 
\citep{2006ApJ...649..599K,2007ApJ...667..176G}. The uncertainties on 
$\theta_E$ were reported to be around 5\%, so we use this value for all 
Einstein radii when calculating the error on $M_E$.
Note that the lensing modeling uses a Singular Isothermal Ellipsoid (SIE) 
mass model, but the resulting projected mass distribution is parameterized 
by an Einstein radius so that the enclosed mass in the projected ellipse is 
the same that would be enclosed in a projected circle from an equivalent 
Singular Isothermal Sphere. This is the radius we adopt here.
Indeed, the Einstein radius determined this way is a robust attribute of 
the lens, being little sensitive to the lens model used (see Kochanek's 
contribution in \cite{2006glsw.conf.....M}).

However the Einstein mass captures not only the lensing effect of the 
lens galaxy, but also the lensing effect of all LOSs over- and underdensities.
We model this LSS contribution by subtracting an effective external lensing 
convergence given by cosmology, $\kappa_{ext}$.
The real lensing mass of the lens galaxy is therefore given by 
\citep{2010ApJ...711..201S}.
\begin{equation}
m_L(\kappa_{ext}) = M_E - \pi R_E^2 \kappa_{ext} \Sigma_{cr}
= (1-\kappa_{ext}) M_E  \; .
\label{lensing_mass}
\end{equation}
Note that the external convergence field can assume positive and negative 
values, therefore it can both decrease or increase the lensing mass.
In general we do not know the value of $\kappa_{ext}$ for a given lens, so 
it has to be treated as a random value drawn from a probability distribution.
That will be further discussed in Section \ref{bayes}.

\subsection{Dynamical Mass}

\begin{figure*}
\epsscale{1.6}
\plotone{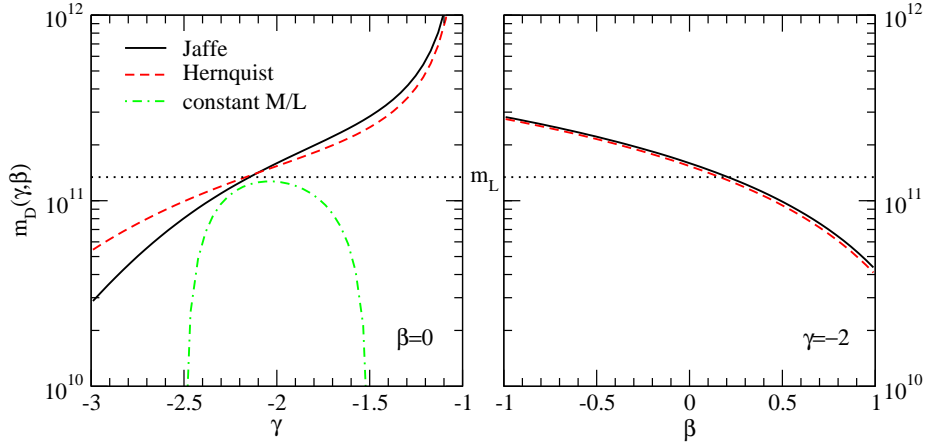}
\vspace{0.4cm}
\caption{Dynamical mass behavior in relation to density profile logarithmic 
slope, $\gamma$, and the velocity anisotropy parameter, $\beta$, in 
units of $h^{-1}M_{\sun}$. The curves were calculated for an ``average'' 
system with $z_l=0.2$, $z_s=0.6$, $\sigma_{ap}=250 kms^{-1}$, 
$\theta_E=1.2\arcsec$, $\theta_{ef}=2.2\arcsec$. 
Solid lines use Jaffe and dashed lines use Hernquist light distribution profiles. 
The dotted line depicts the strong lensing mass, Equation (\ref{M_E}),
for the ``average'' system: $M_E=1.34\cdot 10^{11}h^{-1}M_{\sun}$.
The dot-dashed line on the left panel was obtained using 
$\nu(r)\propto\rho(r)\propto r^{\gamma}$.} 
\label{dynamic_mass_fig}
\vspace{0.4cm}
\end{figure*}

We call the dynamical mass, $m_D$, the mass estimated from the observed 
velocity dispersion. 
Here we are interested in examining the case of a power law for the 
density profile, $\rho = A r^\gamma$, where $A$ is a constant that 
has to be determined from the Jeans equation and the observed velocity 
dispersion.
The mass within the cylinder $C_E$ of Einstein radius $R_E$ is then
\begin{equation}
  m_D(\gamma,\beta) = \int_{C_E} \rho(r) dV = \frac{2\pi^{3/2}}{3+\gamma}
  \frac{\Gamma\left(-\frac{1+\gamma}{2}\right)}
       {\Gamma\left(-\frac{\gamma}{2}\right)} A R_E^{3+\gamma}.
\label{dynMass}
\end{equation}
The spherical Jeans equation \citep{binney87} can be written as
\begin{equation}
  \frac{1}{\nu}\frac{d (\nu \sigma_r^2)}{dr}+2\frac{\beta \sigma_r^2}{r}
  = - \frac{d\Phi}{dr}=-\frac{\pi G}{3+\gamma} A r^{1+\gamma},
  \label{Jeans}
\end{equation}
where $\sigma_r$ is the radial velocity dispersion, 
$\nu(r)$ is the luminosity density profile 
\citep{1983MNRAS.202..995J,1990ApJ...356..359H},
$\beta\equiv1-\sigma_t^2/\sigma_r^2$ is the anisotropy parameter 
of the velocity distribution ($\sigma_t$ is the tangential velocity 
dispersion), and $\Phi$ is the gravitational potential 
produced by the assumed density profile.

Since the quantity observationally available is the luminosity-weighted average 
velocity dispersion within a given aperture, $\sigma^2_{ap}$, 
the following constraint is necessary for the determination of the 
constant $A$:
\begin{equation}
\sigma^2_{ap}= \frac {\int_{C_{ap}} \nu \sigma_r^2 dV}
{\int_{C_{ap}} \nu dV} \; , \label{sigma2}
\end{equation}
where the integration volume is an infinite cylinder of radius $R_{ap}$ 
with axis along the LOS.

To simplify, and since there is very little prior knowledge on the 
velocity anisotropy parameter, we assume that $\beta$ is a constant.
In Appendix \ref{app_jeans}, we give more details on the solution of Jean's 
Equation (\ref{Jeans}), and in Appendix \ref{app_seeing} we examine 
the correction due to seeing effects.

Figure \ref{dynamic_mass_fig} displays the general behavior of the dynamical 
mass as a function of the density profile slope and velocity 
anisotropy parameter. 
We examine $m_D(\gamma,\beta=0 )$ and  $m_D(\gamma=-2,\beta)$. 
Other combinations around these fixed values give qualitatively similar results.
The dotted line depicts the strong-lensing mass for this hypothetical system, 
so it is possible to glimpse from the intersection of the curves the expected 
value of the dynamical parameters, $\gamma$ and $\beta$.
The use of the Jaffe or the Hernquist light profiles do not change the 
qualitative behavior of the curves.
The use of a constant mass to light ratio, 
$\nu(r)\propto\rho(r)\propto r^{\gamma}$, displayed as the dot-dashed 
line in the left panel of Figure \ref{dynamic_mass_fig}, gives a very 
distinct and interesting behavior for $m_D(\gamma)$, 
reproducing a result obtained by \cite{2007arXiv0706.3098G}. 
However, a single power law is not a realistic approximation for the 
light distribution of the lens galaxies in the sample.

\subsection{Statistical Analysis}
\label{bayes}

We want to compare the estimates of lensing and dynamical masses taking
into account the possibility that masses obtained through gravitational
lensing are affected by the LSS, $m_L=m_L(\kappa_{ext})$, and that the 
lens galaxy dynamical mass depends only on its density profile and velocity 
anisotropy, $m_D=m_D(\gamma,\beta)$.
Our models, then, can have up to three free parameters: 
$\kappa_{ext}$, $\gamma$ and $\beta$.
 
To construct a likelihood for the system we define the quantity
\begin{equation}
F=\frac{m_L}{m_D}-\frac{m_D}{m_L}.
\end{equation}
Note that both $m_L$ and $m_D$ refer to the projected mass within the 
Einstein radius.
The likelihood for each lens system is then written as
\begin{equation}
{\cal L}_i = \frac 1 {\sqrt{2\pi} \sigma^2_{F,i}} 
\exp \left[- \frac{(F_t-F_{obs,i})^2}{2\sigma^2_{F,i}} \right] ,
\end{equation}
where $F_{obs}=F_{obs}(\kappa_{ext},\gamma,\beta;data)$ is the measured 
$F$ given the model and the observational data, and $F_t$ is the expected 
value for it, which, in the desired case where both $m_L$ and $m_D$ are 
estimates of the same true galaxy mass, corresponds to $F_t=0$. 
Note that other quantities could be defined to construct the likelihood, 
for example 
$F=m_L/m_D+m_D/m_L$ ($F_t=2$), $F=m_L/m_D$ ($F_t=1$), $F=m_L-m_D$ ($F_t=0$). 
The next to last definition gives a likelihood that is not symmetrical between 
$m_L$ and $m_D$, what is not desirable, and the last example has the 
inconvenience of maximizing the likelihood not only in the desired region 
of the parameter space in which $m_L\sim m_D$, but also in the region where 
both mass estimates are small, what introduces artificial solutions that give
maximum LOS contamination. 
However if $\kappa_{ext}=0$ is fixed, then all definitions for $F$, 
including the last, give very similar results.

The variance in $F$ is estimated as being (an index $i$ is implied in all 
quantities)
\begin{equation}
\sigma^2_F=\frac{\left(m_L^2+m_D^2\right)^2}{m_L^4 m_D^4}
\left(m_L^2\sigma^2_D + m_D^2 \sigma^2_L\right),
\end{equation}
where $\sigma_{\{D,L\}}$ is the uncertainty in $m_{\{D,L\}}$.

\begin{figure}
\epsscale{1.}
\plotone{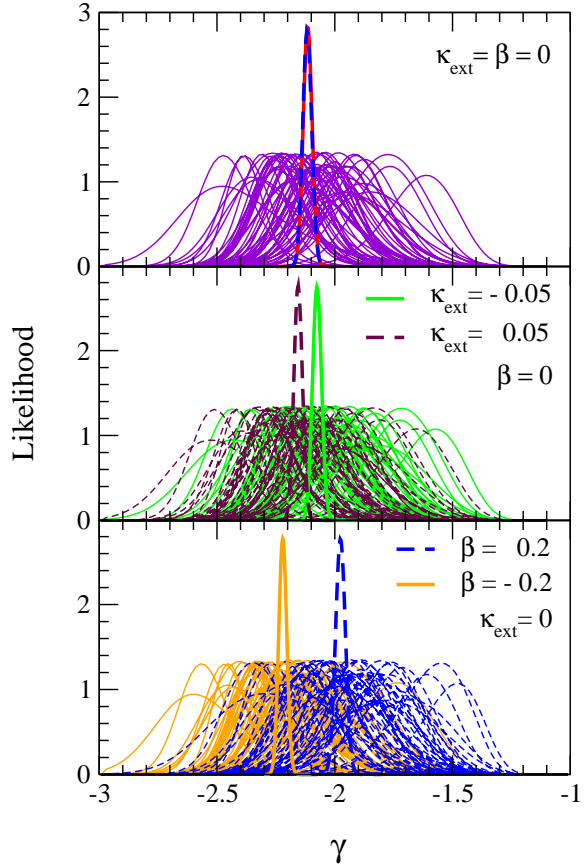}
\vspace{0.cm}
\caption{Likelihood distribution of $\gamma$ values. 
Multiple thin lines are the individual likelihoods for each lens. 
The thick spiky curves are the likelihoods for the whole set 
(rescaled for better visualization).
It was adopted Jaffe's light profile in all panels.
Each panel depicts particular cuts in the parameter space, indicated in the 
internal legend.
In the top panel, the solid line does not take seeing correction into account 
and the dashed line does. }
\label{Li_fig}
\vspace{0.cm}
\end{figure}

The joint likelihood for the whole set of $N$ galaxies is then 
\begin{equation}
{\cal L}(\kappa_{ext},\gamma,\beta) = \prod _{i=1}^N  {\cal L}_i .
\end{equation}
Figure \ref{Li_fig} shows the individual $\gamma$ likelihood distributions 
for each lens, as well as the joint likelihood for the whole sample, 
imposing fixed null external convergence and velocity anisotropy, 
recreating a similar figure obtained by \cite{2009ApJ...703L..51K}.
In the same figure we examine the impact of the seeing correction 
(see Appendix \ref{app_seeing}).
The two narrow and almost indistinguishable curves differ just in that in 
their calculation one takes into account the seeing correction (dashed line) 
and the other (solid line) does not.
The seeing correction is negligible.

The posterior probability distribution of the model parameters given the data 
is determined by the Bayes' theorem
\begin{equation}
P(\kappa_{ext},\gamma,\beta) = \frac{\pi(\kappa_{ext},\gamma,\beta)
{\cal L}(\kappa_{ext},\gamma,\beta)}{E},
\label{posterior}
\end{equation}
where
\begin{equation}
 E \equiv \int \pi(\kappa_{ext},\gamma,\beta)\,{\cal L}(\kappa_{ext},\gamma,\beta)
 \, d\kappa_{ext} d\gamma d\beta 
\end{equation}
is the Bayesian evidence \citep{2008ConPh..49...71T} and 
$\pi(\kappa_{ext},\gamma,\beta) = \pi(\kappa_{ext}) \pi (\gamma) \pi (\beta)$ 
is the product of the prior probability distributions for the model parameters. 

We adopt uniform priors for two of the parameters, 
$\pi(\gamma)=\pi(\beta)=1/2$ for $\gamma \in [-3,-1]$ and for 
$\beta \in [-1,1]$ and $\pi(\gamma)=\pi(\beta)=0$ outside these intervals.
For $\kappa_{ext}$ we assume a prior given by the distribution of expected 
values for the convergence on a random LOS in the concordance cosmology 
adopted here (null mean value). 
We approximate this external convergence probability distribution function 
by the lognormal expression given by 
\cite{2002ApJ...571..638T}, 
$\pi(\kappa_{ext})=P_{ln}(\kappa_{ext})$, with 
$\langle \kappa^2 \rangle ^{1/2}=0.025$ and $\kappa_{min}=-0.06$. 

We estimate the free parameters $p$ of the models using the Bayes estimator 
(also called the posterior mean) 
\begin{equation}
\langle p \rangle \equiv \int {P(p) p \, dp}, 
\label{post_mean}
\end{equation}
and also calculate the root mean square deviation, 
rms$(p)=\sqrt{\langle p^2\rangle -{\langle p\rangle^2}}$,
which give a measure of the parameter value dispersion.
Note that the parameter estimate is made from the whole data set considered
altogether and not for each individual galaxy-lens system. 
The models are meant to be underlying models for all systems, independently 
of the particularities of each lens.
In this approach the information about the sample dispersion among all lenses 
is mixed with the information about individual uncertainty in each lens.

Two models, $M_1$ and $M_2$, even with different degrees of freedom, 
can be compared by the ratio of their Bayesian evidence, 
also known as Bayes factor, 
\begin{equation}
B(M_1,M_2)= \frac {E(M_1)}{E(M_2)} , 
\end{equation}
in which values can be interpreted qualitatively using Jeffrey's scale 
\citep{2008ConPh..49...71T}.
The strength of the evidence in favor of the model with larger $E$ is 
called inconclusive if $\vert \ln B \vert <1$, weak if above this value 
and below $2.5$, moderate if $2.5<\vert \ln B \vert <5$, and strong if 
above that.

The main appeal of this approach for model comparison is that the Bayesian 
evidence automatically implements Occam's razor by penalizing more strongly 
more complex models, those with more free parameters.

We sample the parameter space using a grid finer than the typical 
features scales of the system, which are characterized by the rms of each 
parameter. For example, for models with three free parameters we use as 
grid spacing:
$\delta \kappa_{ext} = 0.0036$, 
$\delta \gamma = 0.018$, 
$\delta \beta = 0.03$;
for a model with only $\gamma$ free we use $\delta \gamma = 0.004$. 
Such grids allow a complete sampling of the relevant parameter space and 
guarantees an appropriate probing of the likelihood and reliable estimates 
of parameter values and Bayesian evidence.

\section{Results}
\label{results}

\begin{table*}
\begin{center}
\caption{\label{models_results}Models' Results Summary. Parameters posterior mean and rms, and Bayesian evidence. Values inside parentheses indicate fixed values in the model.}
\begin{tabular}{lcllllllc}
\tableline\tableline
Free Parameters & Light Profile & $\langle \kappa_{ext} \rangle$ &
 ${\rm rms}(\kappa_{ext})$ &  $\langle \gamma \rangle$ & ${\rm rms}(\gamma)$ & 
 $\langle \beta \rangle$& ${\rm rms}(\beta)$ & $\ln E$ \\
\tableline
0 & Jaffe     & (0)    & ...   & (-2)   & ...   & \,(0)  & ...  & -49.7 \\
1 & Jaffe     & (0)    & ...   & -2.115 & 0.019 & \,(0)  & ...  & -36.1 \\
2 & Jaffe     & -0.005 & 0.022 & -2.111 & 0.026 & \,(0)  & ...  & -36.1 \\
2 & Jaffe     & (0)    & ...   & -2.06  & 0.12  & \,0.07 & 0.19 & -37.3 \\
3 & Jaffe     & -0.003 & 0.023 & -2.06  & 0.12  & \,0.06 & 0.19 & -37.4 \\
3 & Hernquist & -0.004 & 0.022 & -2.16  & 0.11  &  -0.06 & 0.14 & -39.8 \\
\tableline
\end{tabular}
\end{center}
\end{table*}

We have analyzed the data described in Section \ref{data} with models with 
up to three degrees of freedom, one ($\kappa_{ext}$), corresponding to the 
LOS-LSS contamination, that would affect the strong lensing estimate of the 
lens mass and two ($\gamma$ and $\beta$), corresponding to intrinsic 
properties of the lens, that determine its dynamical mass estimate.
Two light distribution profiles (Jaffe and Hernquist) were examined.

Table \ref{models_results} summarizes some results for the models considered.
The simplest model with no free parameters (isothermal density profile, 
no external convergence and velocity anisotropy) has a much lower 
Bayesian evidence than more complex models, so it can be said that it 
is strongly disfavored.
A model with free density profile logarithmic slope has a much higher 
Bayesian evidence and yields a determination of this parameter with 
a 1\% precision, being slightly steeper than an isothermal profile.
This result is similar to what was found by \cite{2009ApJ...703L..51K}.

If instead of estimating $\langle \gamma \rangle$ and ${\rm rms}(\gamma)$ 
from Equation (\ref{post_mean}) we find for each lens galaxy a $\gamma_i$ 
(the point of maximum likelihood for each ${\cal L}_i$) and then proceed the 
calculation of the average value and sample deviation, we find 
$-2.12$ and $0.17$, respectively. 
That is in broad agreement with what is found from the posterior probability 
for the whole data set (\ref{posterior}) (remember that there is a factor 
$\sqrt N$ between the sample and standard deviations before comparing with 
the value in Table \ref{models_results}).
The sample deviation found can be considered a reasonable estimate of the sample dispersion on $\gamma$.\footnote{Note that Equation (7) of \cite{2006ApJ...649..599K} to obtain the intrinsic scatter leads approximately to the sample deviation if 
$\delta \gamma_i \ll \sigma_\gamma$.} 

The addition of a degree of freedom due to an external convergence, 
constrained by a prior probability, generating a model with two degrees 
of freedom, does not alter the mean density profile slope, but increases in 
almost 40\% the uncertainty in its determination.

A model with no external convergence and freedom in the density profile and
anisotropy parameter allows a profile closer to isothermal, and in fact 
compatible with it, given its much larger dispersion on $\gamma$ 
(six times larger than for a model with only $\gamma$ free).
Such model is compatible with isotropy, but is weakly disfavored in 
relation to a model with free density profile and fixed null anisotropy.
We note that \cite{2009ApJ...703L..51K} found a positive anisotropy 
(significantly distinct from isotropy) for the same sample, but using for 
that an independent determination based on scaling relations of the 
density profile logarithmic slope. 
That is a different method from ours, which relies solely on the joint 
strong lensing and dynamical analysis.

A more complex model, with three degrees of freedom, does not give results 
much different from those of the model with free $\gamma$ and $\beta$ and 
fixed $\kappa_{ext}=0$. 
A similar model, but with a different luminosity density profile (Hernquist), 
has a slightly steeper density profile and more negative anisotropy parameter, 
being weakly disfavored in comparison to the model with Jaffe's light profile.

If both external convergence and anisotropy are fixed, 
$\kappa_{ext}=\beta=0$, then $\langle \gamma \rangle=-2.12$ for both light 
profiles and ${\rm rms}(\gamma)=0.019$ (using Jaffe) and 
${\rm rms}(\gamma)=0.027$ (using Hernquist). 
Therefore, the neglect of our ignorance on $\beta$ and, to a lower extent, 
$\kappa_{ext}$, by fixing them equal to zero, implies a considerable 
underestimate of the dispersion in $\gamma$ and possibly the introduction 
of a systematical error, since in the more complex model 
$\gamma =-2.06 \pm 0.12$ (using Jaffe) and 
$\gamma =-2.16 \pm 0.11$ (using Hernquist).

The results for the posterior mean and rms for the external convergence 
suggest that the system likelihood did not alter the prior probability 
distribution for $\kappa_{ext}$, which had $\langle \kappa_{ext} \rangle =0$ and 
$\langle \kappa_{ext}^2 \rangle^{1/2}=0.025$.
The small difference may be attributed to the computational truncation of the 
calculations at $\kappa_{ext}\sim 0.15$, excluding rare events of larger value.

\begin{figure*}
\epsscale{0.7}
\plotone{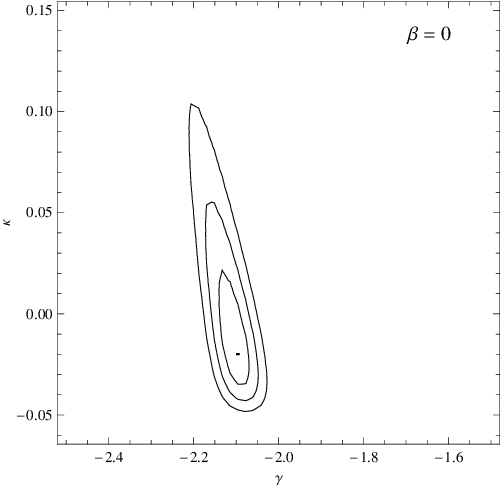}
\plotone{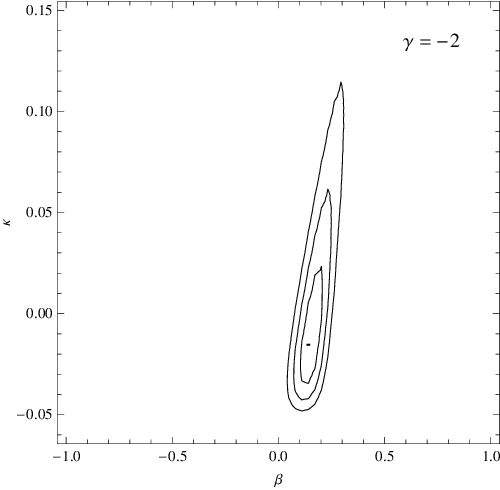}
\plotone{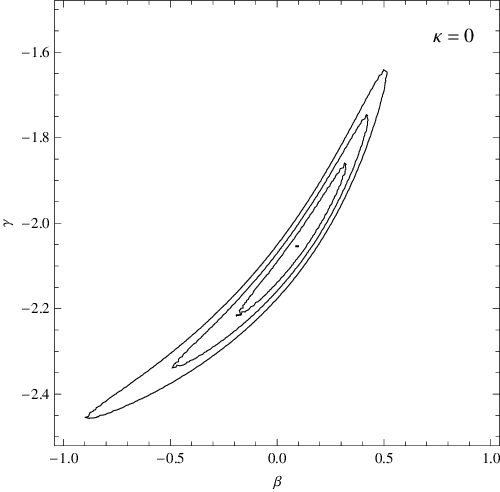}
\\
\plotone{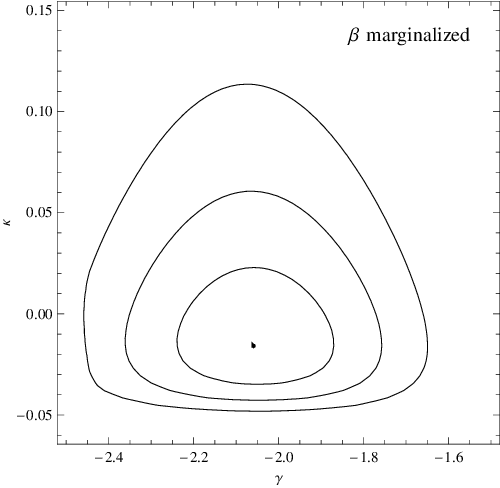}
\plotone{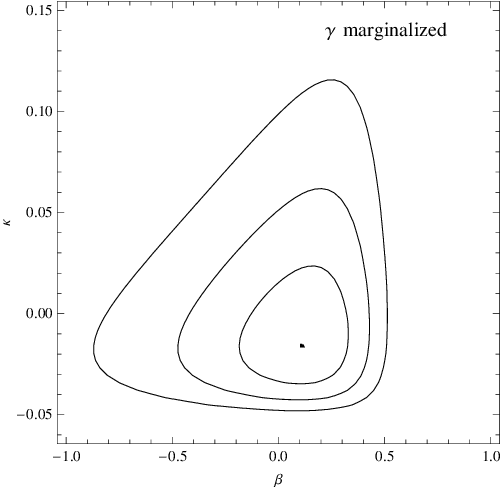}
\plotone{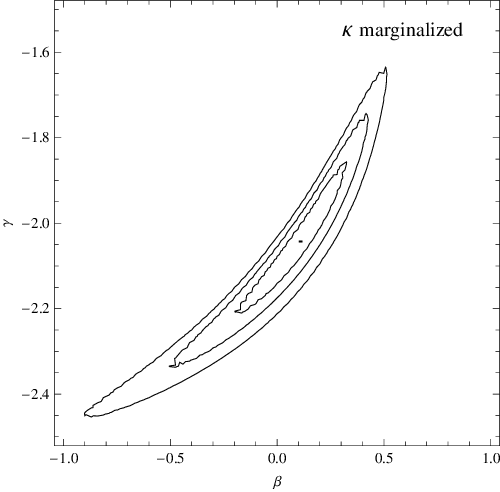}
\caption{Posterior probability distribution. 
Top row has one of the three free parameters fixed.
Bottom row has one of the three free parameters marginalized over.
The point of maximum probability and the $1\sigma$, $2\sigma$ and $3\sigma$ 
confidence levels are shown.
Jaffe's light distribution profile is used.}
\label{fig_likelihoods2D}
\end{figure*}

\begin{figure*}
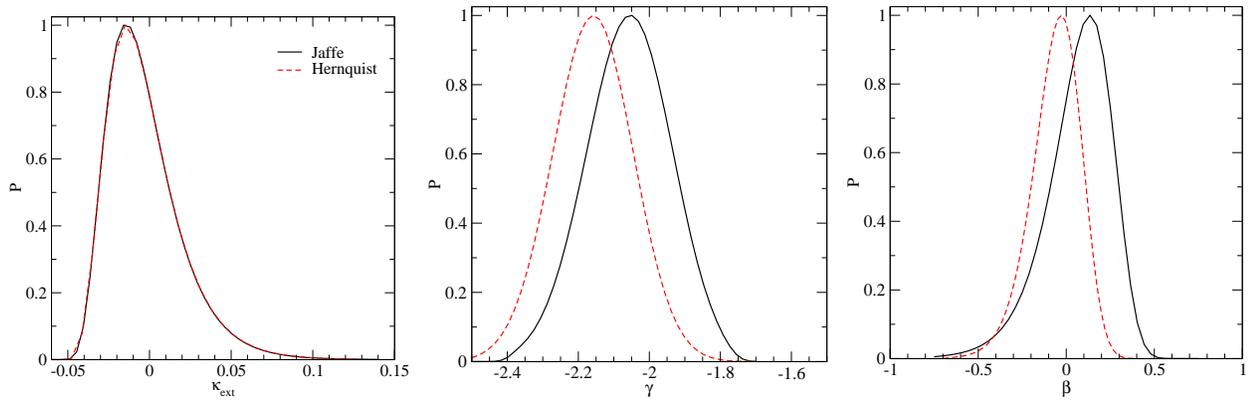

\epsscale{0.7}
\plotone{fig4a.eps}
\plotone{fig4b.eps}
\plotone{fig4c.eps}
\caption{Posterior probability distribution marginalized over two parameters.
Solid lines use Jaffe's light profile and dashed lines use Hernquist's.} 
\label{fig_likelihoods1D}
\end{figure*}

Figure \ref{fig_likelihoods2D} shows posterior probability contours 
in two dimensions (free parameters in a model) for models using Jaffe's 
light profile (the use of Hernquist's would not change the conclusions).
The top row panels have models with two free parameters and one fixed,  
and the bottom panels have the models with three free parameters, but one 
of them marginalized over.
When we compare the top and bottom panels at the first column it shows 
that the prior ignorance on the anisotropy parameter implies a more uncertain 
determination of the density profile.
The second column shows that the reciprocal is also true, and the third 
column clearly depicts this degeneracy between density profile and anisotropy 
parameter, which was already hinted in Figure \ref{dynamic_mass_fig}, 
since the curves for $m_D(\gamma,\beta=0)$ and $m_D(\gamma=-2,\beta)$ 
have monotonically increasing and decreasing behavior, respectively.
The effect of taking into account a prior ignorance on $\kappa_{ext}$ is to 
fatten the banana-shaped contour of the  $\gamma \times \beta$ posterior 
probability plot, but not changing the projected limits on $\gamma$ and 
$\beta$.

In Figure \ref{fig_likelihoods1D} are shown the parameter posterior 
probability distributions marginalized over two parameters for models with 
three degrees of freedom.
Results for both light profiles (Jaffe and Hernquist) are presented, 
being qualitatively the same.
The posterior distribution for the external convergence along the lenses 
LOSs (left panel) is indistinguishable from the prior distribution 
used, which indicates that the likelihood distribution in $\kappa_{ext}$ 
is much broader than the prior used.

The marginalized posterior probability distributions for the density profile 
logarithmic slope (central panel of Figure \ref{fig_likelihoods1D}) 
follow a near Gaussian behavior and illustrates what was already observed 
in Table \ref{models_results}. 
The use of Hernquist's light profile implies a slightly steeper lens 
density profile than Jaffe's profile.

The right panel of Figure \ref{fig_likelihoods1D} shows the marginalized 
posterior probability distributions for the anisotropy parameter. 
It shows that our prior ignorance on the velocity anisotropy, quantified 
as a flat prior in the interval $[-1,1]$, is somewhat reduced, that the 
lenses are on average compatible with isotropy, but that positive and 
negative anisotropies are allowed, and that the use of Jaffe's light profile 
favors slightly more positive values for $\beta$ than Hernquist's.

\section{Discussion and Conclusions}
\label{conclusion}

Two independent methods, strong gravitational lensing and stellar dynamics, 
were used to determine the projected galaxy mass within its Einstein radius 
for a set of 58 galaxies from SLACS.
From the comparison of the two masses, the lens density profile, 
velocity anisotropy and LOS external convergence were probed in an 
integrated Bayesian approach.

The use of a prior probability distribution for the external convergence 
allowed us to quantify the effect of the LSS on the problem of joint lensing 
and dynamical analysis of SLACS lenses.
The data do not seem have enough information to increase our knowledge 
on the external convergence for the particular set of LOS probed by the 
lenses since the posterior probability did not differ from the prior.
The shapes of the $\kappa_{ext} \times \gamma$ and $\kappa_{ext} \times \beta$ 
degeneracies indicate that the determination of the density profile and 
anisotropy parameter are not very sensitive to the value of the LOS 
convergence in the range allowed by the prior probability distribution 
used. 
This can be understood from the functional dependence of the dynamical mass 
with $\gamma$ and $\beta$ (Figure \ref{dynamic_mass_fig}).
A small variation in $\gamma$ or $\beta$ implies a large change in $m_D$.
This may mean that the uncertainties in the system due to the lens structure 
are more determinant than the role played by the external convergence and/or 
that SLACS lenses are in fact an unbiased sample in relation to a random 
LOS, despite the fact that the lenses are elliptical galaxies which are 
often found in dense regions.
That would be in agreement with \cite{2009ApJ...690..670T} who find large 
external convergence values in just few extreme cases and with 
\cite{2005ApJ...635L...1W}, for whom secondary matter along the LOS of 
strong lenses is strongly dependent on source redshift, being rare for sources 
with $z<1$, which is the case for most SLACS lenses.
Another argument for why SLACS would not be biased to be in LOS
with higher than average densities, as would be suggested by some theoretical 
works \citep{2007MNRAS.382..121H}, is that they were spectroscopically 
selected. 
In contrast, lenses selected by photometric image identification would be 
more prone to be biased toward higher than average overdense LOS because 
lensed images of larger separation angle, tracing the densest regions 
\citep{2009ApJ...695.1233F}, would be more easily identified, which does not 
happen in a spectroscopically selected sample.

The joint strong lensing and stellar dynamics analysis does improve our 
prior knowledge on the anisotropy parameter, but does not strongly constrain 
it, its probability distribution being broad and statistically compatible 
with isotropy on average.
Nevertheless, the degeneracy between $\gamma$ and $\beta$ gives an 
indeterminacy that correlates a larger anisotropy with a flatter 
density profile, which can also be understood from the functional behavior 
of $m_D(\gamma,\beta)$ (Figure \ref{dynamic_mass_fig}).
An increment in $\gamma$ can be annulled, in terms of a variation in the 
dynamical mass, by a decrement in $\beta$, and vice versa.

The inclusion of two degrees of freedom in the model with respect to the 
LSS convergence and velocity anisotropy allows us to take into 
account and examine the effect of our prior ignorance on these features 
of the lens system. 
The most visible effect of $\kappa_{ext}$ and $\beta$ on the determination 
of the density profile logarithmic slope is the considerable broadening of its 
probability distribution, which means a more uncertain determination of 
$\gamma$ than what is suggested by a simpler model that does not take into 
account those features.
Within the uncertainty found in the more complex model, the density profile 
is statistically compatible with an isothermal profile.
This very particular density profile is also found by other authors, 
apparently as a result of the complementarity of baryonic and dark matter 
profiles \citep{2005astro.ph..7056H,2005ApJ...623L...5F,2006ApJ...648..826L,
2009JCAP...01..015B,2008MNRAS.384..987C}. 

We used standard assumptions and approximations in our modeling of galaxies 
and analysis.
Nevertheless, we can identify several areas where further work can be done to 
refine the understanding of these strong-lensing systems, which can also 
be seen as caveats to the present works in the area.
Among them we highlight 
(1) the triaxiality and substructure of lens halos, whose importance was already suggested by \cite{2005astro.ph..9323M} and 
\cite{2006ApJ...643..154Y} in the context of simulations on cluster scales, 
(2) the correction of the dynamical mass estimate due to rotational support.
It is well known  that some early-type galaxies can have a significant 
rotational component \citep[e.g.,][]{2007MNRAS.379..401E} and two-dimensional 
kinematics for some few SLACS lenses are already becoming available 
\citep{2008MNRAS.384..987C,2009MNRAS.399...21B},
(3) the brightness distribution, which could be treated more 
realistically with the observed full surface luminosities for the 
individual lenses, instead of individualized fits of a universal profile, and 
(4) the velocity anisotropy parameter, which was assumed to be a 
constant, but that more realistically must be a function of radius. 
However, very little is known about the velocity anisotropy of early-type 
galaxies, be it observationally, theoretically, and even from simulations.

The relaxation of some of our assumptions, with the almost inevitable 
addition of extra free parameters, could prove a fruitful source of 
investigation; however, it would likely require a larger galaxy sample to 
reduce the likely increased degeneracies among the degrees of freedom 
of the model.
As we have illustrated, models with a large number of parameters may have a 
higher likelihood but lower Bayesian evidence, since the added complexity 
must pay its price in a Bayesian sense.

\acknowledgments
The authors thank CNPq and FAPESP for financial support, the SLACS and SDSS 
teams for the databases used in this work, and the referee whose comments 
helped to clarify many aspects of this paper.

\appendix
\twocolumn
\setcounter{section}{1}
\section*{Appendix}

\subsection{Solving Jean's Equation}
\label{app_jeans}

The spherical Jeans Equation (\ref{Jeans}) can be rewritten, defining 
$x\equiv r/R_{ef}$ (dimensionless) and 
\begin{equation}
  y\equiv \frac {3+\gamma}{4\pi GA} R_{ef}^{-(2+\gamma)} (\nu \sigma_r^2),
\end{equation}
as
\begin{equation}
  \frac{dy}{dx}=-2\beta \frac{y}{x}-\nu x^{1+\gamma}.
  \label{Jeans_new}
\end{equation}

The luminosity distribution is well approximated by the profiles 
\citep{2006ApJ...649..599K}
\begin{equation}
\nu(x) \propto \frac{1}{x^{\gamma_\ast}(x+x_\ast)^{4-\gamma_\ast}} \; , 
\label{HJlight} 
\end{equation}
where $\gamma_\ast=1$ and $x_\ast=1/1.8153$ \citep{1990ApJ...356..359H}, 
or $\gamma_\ast=2$ and $x_\ast=1/0.7447$ \citep{1983MNRAS.202..995J}.
We examine both profiles.

The first-order linear differential Equation (\ref{Jeans_new}) has a solution
\begin{equation}
  y(x) = x^{-2\beta} \left[ C - \int \nu(x) x^{1+\gamma+2\beta}dx \right],
  \label{Jeans_solution}
\end{equation}
where $C$ is an arbitrary constant and the most evident boundary condition 
is $y(x\rightarrow \infty)=0$.
The analytical solution for the integral in Equation (\ref{Jeans_solution}) 
with Equation (\ref{HJlight}) for $\nu(r)$ has the hypergeometric function 
$_2F_1$, which has a computationally demanding solution.
Therefore, we solve Equation (\ref{Jeans_new}) using a fourth-order Runge-Kutha 
algorithm, starting at $y(x=1000)=0$ and evolving $y(x)$ down to $x=0.01$.

The dynamical mass (Equation (\ref{dynMass})) within the Einstein radius $R_E$ is then
\begin{equation}
  m_D(\gamma,\beta) =  \frac{\pi^{1/2}}{2G}
  \frac{\Gamma\left(-\frac{1+\gamma}{2}\right)}
       {\Gamma\left(-\frac{\gamma}{2}\right)} 
  \sigma_{ap}^2 R_E \left(\frac{R_E}{R_{ef}}\right)^{2+\gamma}
  \frac{\int_{C_E} \nu dV}{\int_{C_E} y dV},
\end{equation}
and its error is estimated from the observational errors on 
$\sigma_{ap}$ and $R_E$ through error propagation.

\subsection{Seeing}
\label{app_seeing}
We model the effect of the seeing through a Gaussian smoothing of the 
galaxy projected luminosity. Therefore, the observed surface brightness 
profile is related to an intrinsic (no seeing) profile by
\begin{equation}
I_{obs}(\theta)= \frac{e^{-\theta^2/2\sigma_s^2}}{\sigma_s^2}
\int_0^\infty{I(\theta^\prime){\rm I_0}
\left(\frac{\theta \theta^\prime}{\sigma_s^2}\right)
e^{-\theta^{\prime 2}/2\sigma_s^2} \theta^\prime d\theta^\prime} \; ,
\label{suface_bright}
\end{equation}
where ${\rm I_0}$ is the modified Bessel function of first kind, 
and $\sigma_s^2$ is the Gaussian seeing variance. 
We use $\sigma_s=0.64 \arcsec$, which corresponds to an FWHM of $1.5\arcsec$. 

The seeing correction of the average velocity dispersion within the 
observational aperture is then given by
\begin{equation}
\frac{\sigma_{ap}^2}{\left(\sigma_{ap}^2\right)_{obs}} =
\frac{\int_0^{R_{ap}} \sigma_p^2(R)I(R)R dR}
{\int_0^{R_{ap}} \left[ \sigma_p^2(R)I(R) \right]_{obs}
R dR}
\frac{\int_0^{R_{ap}} I_{obs}(R)R dR}
{\int_0^{R_{ap}} I(R)R dR} \; ,
\label{seeing_correct}
\end{equation}
where $\sigma_p^2$ is the projected velocity dispersion profile,  
$\left[ \sigma_p^2(R)I(R) \right]_{obs}$ is defined 
in an analogous way to $I_{obs}(R)$ in Equation (\ref{suface_bright}),
and the projection is calculated through \citep{binney87}
\begin{equation}
I(R)\sigma_p^2(R)=2\int_R^\infty \left( 1-\beta\frac{R^2}{r^2}\right)
\frac{\nu\sigma_r^2r}{\sqrt{r^2-R^2}}dr .
\end{equation}


\end{document}